 \documentclass{emulateapj}

\usepackage{apjfonts}

\newcommand{\object}[1]{#1}

\slugcomment{to appear in the Astrophysical Journal, Letters}

\shorttitle{Supersoft X-ray Phase of RS Oph}
\shortauthors{Hachisu and Kato}

\begin{document}


\title{Prediction of Supersoft X-ray Phase during the 2006 Outburst
of RS Ophiuchi}


\author{Izumi Hachisu}
\affil{Department of Earth Science and Astronomy,
College of Arts and Sciences,
University of Tokyo, Komaba 3-8-1, Meguro-ku, Tokyo 153-8902, Japan}
\email{hachisu@chianti.c.u-tokyo.ac.jp}

\and

\author{Mariko Kato}
\affil{Department of Astronomy, Keio University, 
Hiyoshi 4-1-1, Kouhoku-ku, Yokohama 223-8521, Japan}
\email{mariko@educ.cc.keio.ac.jp}


%\altaffiltext{1}{Visiting Astronomer, Cerro Tololo Inter-American Observatory.
%CTIO is operated by AURA, Inc.\ under contract to the National Science
%\altaffiltext{2}{Society of Fellows, Harvard University.}
%\altaffiltext{3}{present address: Center for Astrophysics,
%    60 Garden Street, Cambridge, MA 02138}
%\altaffiltext{4}{Visiting Programmer, Space Telescope Science Institute}
%\altaffiltext{5}{Patron, Alonso's Bar and Grill}


\begin{abstract}
\object{RS Ophiuchi} is one of the well-observed recurrent novae
and also a candidate progenitor of Type Ia supernovae.
Its sixth recorded outburst was discovered on 12 February 2006.
Detection of a supersoft X-ray phase will provide a firm 
confirmation of hydrogen shell-burning on the white dwarf and
its turn-on/turn-off dates strongly constrain a mass range of
the white dwarf, which clarify whether or
not \object{RS Oph} becomes a Type Ia supernova.
For a timely detection of its supersoft X-ray phase, 
we have calculated outburst evolution of \object{RS Oph} 
based on the optically thick wind theory and
predicted a supersoft X-ray phase for the 2006 outburst:
it will most probably start on day $80-90$ and continue until
day $110-120$ after the optical maximum.  Its duration is so short
as only a month or so.  We strongly recommend multiple observations
during April, May, and June of 2006 to detect turn-on and turn-off
times of the supersoft X-ray phase.
\end{abstract}





\section{Introduction}
     The recurrent nova \object{RS Ophiuchi} has undergone its sixth
recorded outburst on 12 February 2006 \citep{nar06}.
The previous five outbursts occurred in 1898, 1933, 1958, 1967,
and 1985.  One outburst may have been missed between 1898 and 1933
because \object{RS Oph} was not recognized as a recurrent nova.
These $10-20$ yr recurrence periods indicate both that
the white dwarf (WD) mass is very close to the Chandrasekhar mass
and that its mass accretion rate is as large as $\dot M_{\rm acc} 
\sim 1 \times 10^{-7} M_\sun$~yr$^{-1}$ 
\citep[see, e.g., Fig.2 of][]{hac01kb}.
If the WD mass increases after every outburst, \object{RS Oph}
will soon explode as a Type Ia supernova 
\citep[e.g.,][]{nom82, hkn99, hac01kb}.  It is, therefore, 
crucially important to know how close the WD mass is 
to the Chandrasekhar mass and how much mass is left on the WD
after one cycle of nova outburst.  We are able to constrain a mass
range of the WD if turn-on/turn-off of supersoft X-ray
are detected, because it indicates the durations of 
wind mass loss (how much mass is ejected)
and hydrogen shell-burning without wind mass loss (how much mass
is left).  In this Letter, we have calculated supersoft X-ray phases
for \object{RS Oph} and predict turn-on and turn-off dates for its 
timely detection during the current outburst.

     In \S \ref{opticall_thick_wind_model}, we briefly describe our
optically thick wind model for nova outbursts.  The numerical results
and predictions are shown in \S \ref{prediction_of_SSS}.  Discussions
follow in \S \ref{discussion}.

\section{Optically Thick Wind Model} \label{opticall_thick_wind_model}

%% In a manner similar to \objectname authors can provide links to dataset
%% hosted at participating data centers via the \dataset{} command.  The
%% second curly bracket argument is printed in the text while the first
%% parentheses argument serves as the valid data set identifier.  Large
%% lists of data set are best provided in a table (see Table 3 for an example).
%% Valid data set identifiers should be obtained from the data center that
%% is currently hosting the data.
%%
%% Note that AASTeX interprets everything between the curly braces in the 
%% macro as regular text, so any special characters, e.g. "#" or "_," must be 
%% preceded by a backslash. Otherwise, you will get a LaTeX error when you 
%% compile your manuscript.  Special characters do not 
%% need to be escaped in the optional, square-bracket argument.

     After a thermonuclear runaway sets in on a mass-accreting WD,
its photosphere expands greatly to $R_{\rm ph} \gtrsim 100 ~R_\sun$
and the WD envelope soon settles in a steady-state.
The decay phase of novae can be
followed by a sequence of steady state solutions \citep[e.g.,][]{kat94h}.
We have calculated light curves for the 2006 outburst of \object{RS Oph},
using the same method and numerical techniques as in \citet{kat94h},

     We solve a set of equations, i.e., the continuity, equation
of motion, radiative diffusion, and conservation of energy,
from the bottom of the hydrogen-rich envelope through the photosphere.
The winds are accelerated deep inside
the photosphere so that they are called ``optically thick winds.''
We have used updated OPAL opacities \citep{igl96}.
We simply assume that photons are emitted at the photosphere
as a blackbody with the photospheric temperature of $T_{\rm ph}$.
Physical properties of these wind solutions have already been
published \citep[e.g.,][]{hac01ka, hac01kb, hac04k, hkn96,
hkn99, hknu99, hkkm00, hac03a, kat83, kat97, kat99}.
We have integrated 
\begin{equation}
{{d} \over {d t}} \Delta M_{\rm env} = \dot M_{\rm acc}
- \dot M_{\rm wind} - \dot M_{\rm nuc},
\label{basic_eq_wind}
\end{equation}
and followed time-evolution of a nova.  Here, $\Delta M_{\rm env}$ is
the hydrogen-rich envelope mass, $\dot M_{\rm wind}$ the wind mass
loss rate, $\dot M_{\rm nuc}$ the consumption rate of the envelope
mass by nuclear burning, and $\dot M_{\rm acc}$ the mass accretion
rate.  Both the wind mass loss ($\dot M_{\rm wind}$) and
nuclear burning ($\dot M_{\rm nuc}$) rates are calculated from
our wind solution having the envelope mass of $\Delta M_{\rm env}$.
We assume no mass accretion ($\dot M_{\rm acc}= 0$) during the outburst.

     Optically thick winds stop ($\dot M_{\rm wind}=0$)
after a large part of the envelope is blown in the winds.
The envelope settles into a hydrostatic
equilibrium where its mass is decreasing only by nuclear burning.

When the nuclear burning decays, the WD enters a cooling phase, in
which the luminosity is supplied with heat flow from the ash of
hydrogen burning.

     In the optically thick wind model, a large part of
the envelope is ejected continuously for a relatively
long period \citep[e.g.,][]{kat94h}.  After the maximum expansion
of the photosphere, it gradually shrinks keeping the total luminosity
($L_{\rm ph}$) almost constant.
The photospheric temperature ($T_{\rm ph}$) increases with time
because of $L_{\rm ph} = 4 \pi R_{\rm ph}^2 \sigma T_{\rm ph}^4$.
The main emitting wavelength of radiation moves from optical
through supersoft X-ray.
This causes the decrease in optical luminosity
and the increase in UV.  Then the UV flux reaches a maximum.
Finally the supersoft X-ray flux increases after the UV flux decays.
Thus, we can follow the development of supersoft X-ray light
curves \citep[e.g.,][]{hac05k}.

These timescales depend very sensitively on the WD mass if 
the WD mass is very close to the Chandrasekhar mass.  This is because
the WD radius is very sensitive to the increase in mass near
the Chandrasekhar mass.  The timescale also depends weakly on 
the chemical composition of envelopes.

%% In this section, we use  the \subsection command to set off
%% a subsection.  \footnote is used to insert a footnote to the text.


%% This section also includes several of the displayed math environments
%% mentioned in the Author Guide.

\section{Prediction of Supersoft X-ray Phase}
\label{prediction_of_SSS}
     Based on the observation of previous outbursts of \object{RS Oph}, 
\citet{hac00k, hac01kb} derived various physical parameters of
the star from visual and UV light curve fittings.
They concluded that the WD mass is (1) $1.35 \pm 0.01 ~M_\sun$
for the solar composition or (2) $1.377 \pm 0.01 ~M_\sun$
for a low metallicity of $Z=0.004$ and $X=0.7$.
Therefore, we have calculated two cases
for the metallicity, i.e., $Z=0.02$ and $Z=0.004$.

Once the unstable nuclear burning
sets in, convection widely develops and mixes processed helium with
unburned hydrogen.  This process reduces the hydrogen content
in the envelope by 10$-$20\% for massive WDs like in \object{RS Oph}.
We calculated two cases for the hydrogen content, i.e., 
$X=0.5$ (mixing) and $X=0.7$ (no mixing).  To summarize,
we have calculated total of 14 light curves as 
tabulated in Table \ref{turn_on_off}.

%Table 1
%\placetable{turn_on_off}

\begin{deluxetable}{lllrr}
\tabletypesize{\scriptsize}
%%%%\rotate
\tablecaption{Supersoft X-ray turn-on/turn-off time \label{turn_on_off}}
\tablewidth{0pt}
\tablehead{
\colhead{$X$} & \colhead{$Z$} & \colhead{$M_{\rm WD}$} & \colhead{turn 
on\tablenotemark{a}} 
& \colhead{turn off\tablenotemark{a}} \\
\colhead{} & \colhead{} & \colhead{$(M_{\sun})$} & \colhead{(days)} 
& \colhead{(days)}
}
\startdata
0.7 & 0.02 & 1.37 & 56 &  73 \\
    &      & 1.35 & 78 & 113 \\
    &      & 1.34 & 89 & 134 \\
    &      & 1.33 & 101 & 161 \\
    &      & 1.3  & 141 & 248 \\
0.5 & 0.02 & 1.37 &  43 &  53 \\
    &      & 1.35 &  66 & 85 \\
    &      & 1.34 &  70 & 95 \\
    &      & 1.33 &  87 & 120 \\
    &      & 1.3  &  116 & 175 \\
0.7 & 0.004 & 1.377 & 76 & 99 \\
    &      & 1.35  &  124  & 183 \\
0.5 & 0.004 & 1.377 & 63 &  76 \\
    &      & 1.36 &  86 & 109 \\
\enddata
%% Text for table notes should follow after the \enddata but before
%% the \end{deluxetable}. Make sure there is at least one \tablenotemark
%% in the table for each \tablenotetext.
%%\tablecomments{Table \ref{tbl-1} is published in its entirety in the 
%%electronic edition of the {\it Astrophysical Journal}.  A portion is 
%%shown here for guidance regarding its form and content.}
\tablenotetext{a}{days from the optical maximum}
%%\tablenotetext{b}{Another sample footnote for table~\ref{tbl-1}}
\end{deluxetable}

\begin{figure}
\epsscale{1.15}
\plotone{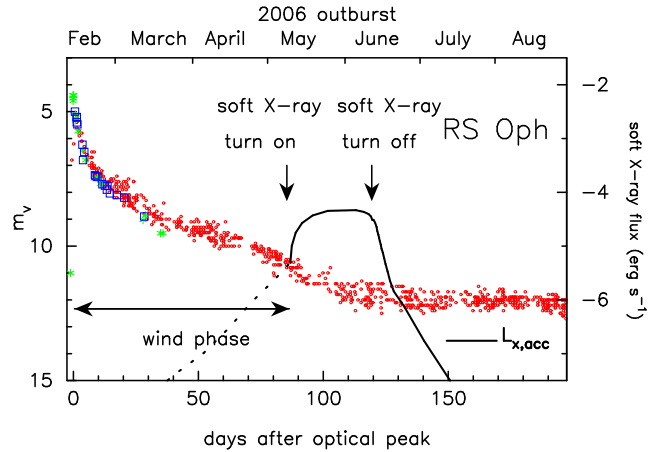}
%\plotone{f1bw.epsi}
%%%\plotone{XpredictM133.epsi}
\caption{
Expected supersoft X-ray fluxes are plotted against time
for our recommended $1.33~M_\sun$ WD model with 
a chemical composition of $X=0.5$ and $Z=0.02$.
We assume the response ($\sim 0.1-10$~keV) of XRT onboard {\it Swift}
(http://swift.gsfc.nasa.gov/docs/swift/),
a distance of $d= 1.0$~kpc, and no interstellar absorption.
We expect a similar response for EPIC CCD MOS 
and a 7 times larger response for EPIC CCD pn both onboard {\it XMM-Newton}.
The optically thick wind continues until day 87 and the hydrogen 
shell-burning ends on day 120.  Visual magnitudes
of the 2006 outburst are taken from VSOLJ ({\it asterisks})
and from IAU Circulars, No.8671, No.8673, No.8681, and No.8688
({\it open squares}).
Visual data of the previous 1985 outburst are taken from AAVSO
({\it small circles}).  X-ray flux by accretion given
in eq. (\ref{accretion_X_eq}) is indicated by a short horizontal
line (labeled by $L_{\rm x, acc}$).
\label{Xpredict}
}
\end{figure}

Figure \ref{Xpredict} depicts our recommended model of \object{RS Oph}.  
Here we assume the response ($\sim 0.1-10$~keV) of XRT
onboard {\it Swift}, a WD mass of $M_{\rm WD}= 1.33~M_{\sun}$, 
a chemical composition of $X=0.5$ and $Z=0.02$, a distance of $d= 1.0$~kpc,
and no interstellar absorption.  We expect similar flux for
EPIC CCD MOS and 7 times larger flux having a similar shape of 
light curve profile for a wider effective area
of EPIC CCD pn both onboard {\it XMM-Newton}.

After the optical maximum, the photosphere gradually shrinks with the
envelope mass being lost in the wind.  The temperature increases with
time but is still lower than $3.5 \times 10^5$~K ($\sim 30$~eV) until
the wind stops on day 87.  Thus, in this wind phase, the expected
supersoft X-ray flux is not so high
as shown in Figure \ref{Xpredict} (dotted part).
Moreover, supersoft X-rays are probably obscured by the optically
thick wind itself because the wind mass loss rate is as large as
$10^{-7} - 10^{-6} ~M_\sun$~yr$^{-1}$.  Supersoft X-rays
are not observed or its flux is very low during the wind phase.
Therefore, we define turn-on time of supersoft X-ray as the time
when the wind stops.  It is 87 days after the optical maximum
in the case of Figure \ref{Xpredict}.
On that day the photospheric temperature is as low as
$\log T_{\rm ph}$~(K)$= 5.55$ but quickly increases to reach 
the maximum temperature of $\log T_{\rm ph}$~(K)$=6.07$ on day 116.

Stable hydrogen shell-burning ends on day 120, where the photospheric
temperature is $\log T_{\rm ph}$~(K)$= 6.05$.
After that, the supersoft X-ray flux as well as the total luminosity
quickly decays.  Then, we define the date of "supersoft X-ray turn-off"
as the time when stable nuclear burning ends \citep[see, e.g.,][]{hac05k}. 

%Fig.1
%\placefigure{Xpredict}

\begin{figure}
%%%\plottwo{Xpredict.epsi}{Xpredict_color.epsi}
\epsscale{1.15}
\plotone{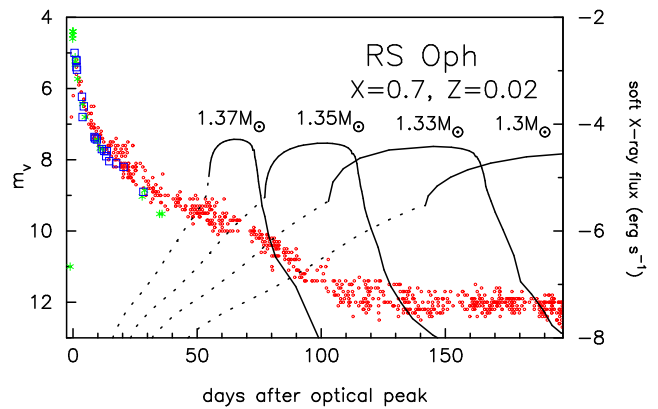}
%\plotone{f2bw.epsi}
%%%\plotone{lightXrayX70Z02.epsi}
\caption{
Same as those in Fig. \ref{Xpredict}, but for the solar composition 
($X=0.7$ and $Z=0.02$) of the envelope.
WD mass is attached to each curve.
\label{lightXrayX70Z02}
}
\end{figure}

Figure \ref{lightXrayX70Z02} shows supersoft X-ray fluxes 
for the WD masses of $M_{\rm WD}= 1.37$, 1.35, 1.33 and 
$1.3~M_{\sun}$.  Here, we assume $X=0.7$ and $Z=0.02$.
The supersoft X-ray phase lasts 107 days for $M_{\rm WD}= 1.3~M_{\sun}$
but only 17 days for $M_{\rm WD}=1.37~M_{\sun}$.

%Fig.2
%\placefigure{lightXrayX70Z02}

\begin{figure}
%%%\plottwo{Xpredict.epsi}{Xpredict_color.epsi}
\epsscale{1.15}
\plotone{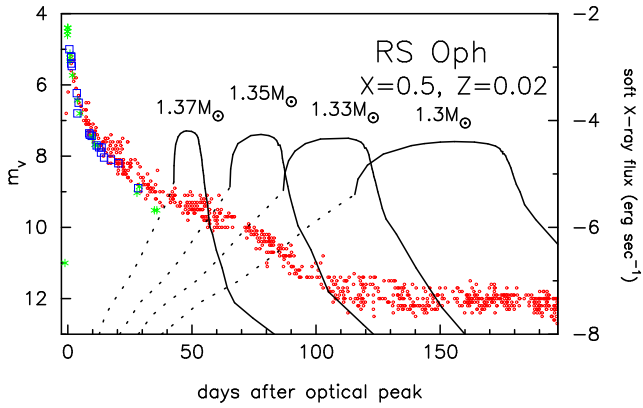}
%\plotone{f3bw.epsi}
%%%\plotone{lightXrayX50Z02.epsi}
\caption{
Same as those in Fig. \ref{lightXrayX70Z02}, but for $X=0.5$ and $Z=0.02$.
\label{lightXrayX50Z02}
}
\end{figure}

Figure \ref{lightXrayX50Z02} shows supersoft X-ray fluxes for 
lower hydrogen content of $X=0.5$ and $Z=0.02$.
This figure shows four cases of the WD mass.
The duration of supersoft X-ray phase becomes shorter than 
those in Figure \ref{lightXrayX70Z02}.

%Fig.3
%\placefigure{lightXrayX50Z02}

\begin{figure}
%%%\plottwo{Xpredict.epsi}{Xpredict_color.epsi}
\epsscale{1.15}
\plotone{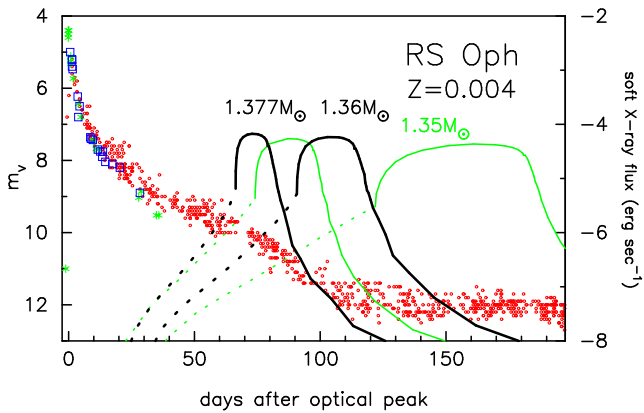}
%\plotone{f4bw.epsi}
%%%\plotone{lightXZ004.epsi}
\caption{
Same as those in Fig. \ref{lightXrayX70Z02}, but for $Z=0.004$
and $X=0.5$ ({\it thick solid}: $1.377$ and $1.36~M_\sun$),
or for $Z=0.004$ and $X=0.7$ ({\it thin solid}: $1.377$ and 
$1.35 ~M_\sun$).
\label{lightX0507Z004}
}
\end{figure}

Low metallicities have been suggested for the giant companion of 
\object{RS Oph} \citep{sco94, con95, smi96}.
Figure \ref{lightX0507Z004} shows supersoft X-ray fluxes 
for a low metallicity of $Z=0.004$.

%Fig.4
%\placefigure{lightX0507Z004}

The development of outburst depends both on the WD mass and on the 
chemical composition.  (1) Less massive WDs evolve more slowly
because of a large envelope mass on the WD.
(2) Less hydrogen contents of envelopes make nova
evolution more quickly because of earlier consumption of hydrogen.
(3) Lower metallicities of envelopes slow down
nova evolution because of less amount of catalyst in the CNO-cycle
and less acceleration of the winds (optically thick winds are
driven by an opacity peak that is due to iron lines).

%% The \notetoeditor{TEXT} command allows the author to communicate
%% information to the copy editor.  This information will appear as a
%% footnote on the printed copy for the manuscript style file.  Nothing will
%% appear on the printed copy if the preprint or
%% preprint2 style files are used.

%% The eqnarray environment produces multi-line display math. The end of
%% each line is marked with a \\. Lines will be numbered unless the \\
%% is preceded by a \nonumber command.
%% Alignment points are marked by ampersands (&). There should be two
%% ampersands (&) per line.

%% Putting eqnarrays or equations inside the mathletters environment groups
%% the enclosed equations by letter. For instance, the eqnarray below, instead
%% of being numbered, say, (4) and (5), would be numbered (4a) and (4b).
%% LaTeX the paper and look at the output to see the results.

%% This section contains more display math examples, including unnumbered
%% equations (displaymath environment). The last paragraph includes some
%% examples of in-line math featuring a couple of the AASTeX symbol macros.

\section{Discussion}
\label{discussion}

%% The displaymath environment will produce the same sort of equation as
%% the equation environment, except that the equation will not be numbered
%% by LaTeX.

In the present 2006 outburst, strong X-rays were already detected
3 days after the discovery with XRT onboard {\it Swift} \citep{bod06}.
Mason et al. fitted the X-ray spectrum with a high-temperature thermal
bremsstrahlung model.  \citet{nes06} reported high-resolution
X-ray spectra of the 2006 outburst, obtained on Feb. 26 (two weeks
after the outburst).  Their {\it Chandra} X-ray observation
indicates high plasma temperatures of $T= (3-60) \times 10^6$~K.

In the 1985 outburst, \object{RS Oph} became a strong X-ray source
with characteristic temperature of a few times $10^6$ K \citep{mas87}.
Mason et al. 
%observed X-rays from 55 days after the
%optical maximum through 251 days with the medium-energy proportional
%counter array (ME) and the low-energy imaging telescope (LE)
%onboard {\it EXOSAT}.
made six separate observations with {\it EXOSAT},
55, 62, 74, 83, 93, and 251 days after the optical maximum.
The source became very weak on and after day 83 in the medium-energy
proportional counter array (ME) and only upper limits were derived.
They interpreted these early X-ray fluxes in terms of
emissions from the pre-outburst M-giant's wind that had been shock-heated
by the outburst ejecta \citep*[see also][]{bod85, ito90, obr92}.

In the later stage, there was still some residual X-ray
emission detected only with the low-energy imaging telescope (LE)
onboard {\it EXOSAT} even on day 251.  \citet{mas87} interpreted
this late X-ray emission in terms of hydrogen shell-burning that still
continues on top of the WD.  This is not consistent
with our model calculation because the steady hydrogen shell-burning
ended long before day 251, i.e., on day $\sim 110-120$
\citep[see][for more detailed discussion]{hac01kb}.  Instead,
\citet{hac01kb} assumed an accretion disk remaining even 
during the outburst and interpreted this soft X-ray emission as accretion
luminosity rather than nuclear burning luminosity, mainly because
the observed flux is too low to be compatible with that for hydrogen
burning.

The accretion luminosity is estimated as
\begin{equation}
L_{\rm x,acc} = {1 \over 2} {{G M_{\rm WD} \dot M_{\rm acc}} 
\over R_{\rm WD}} \sim 3 \times 10^{36} {\rm ~erg~s}^{-1},
\label{accretion_X_eq}
\end{equation}
for an appropriate mass accretion rate of
$\dot M_{\rm acc} \sim 1 \times 10^{-7} M_\sun$~yr$^{-1}$ \citep{hac01kb}.
This accretion X-ray flux is $\sim 300$ times smaller than that for
nuclear burning as indicated in Figure \ref{Xpredict}.

If X-ray satellites will frequently observe \object{RS Oph},
we are able to distinguish X-ray of nuclear burning from that of accretion.
It is because,as shown in Figures \ref{Xpredict}$-$\ref{lightX0507Z004},
we can recognize from the shape of light curves whether it is originated
from nuclear burning or not.  The light curve shape itself is independent
of the distance or absorption.

%Hjellming et al. 1986, ---> 2.4+-0.6E21 hydrogen atom column density
%from radio obs.
We have assumed no interstellar (or circum-binary matter) absorption
in our predicted X-ray fluxes.  Here we discuss absorption effects due
to interstellar or circum-binary matter.  \citet{mas87} referred to a
column density of $\sim (3-4) \times 10^{21}$~cm$^{-2}$ 
toward \object{RS Oph} \citep[see also][]{hje86}.
Although we do not present here how much the flux is reduced due to
absorption by this amount of column density,
it is worth noting that a luminous supersoft
X-ray phase was detected with {\it BeppoSAX} in the 1999 outburst
of the recurrent nova \object{U Sco} \citep{kah99}. 
Kahabka et al. estimated a column density of 
$(1.8-2.6) \times 10^{21}$~cm$^{-2}$ for a blackbody model, or
$(3.1-4.8) \times 10^{21}$~cm$^{-2}$ for a WD atmosphere model.
The WD mass was estimated to be $1.37 \pm 0.01 ~M_\sun$ by \citet{hkkm00},
which is similar to that for \object{RS Oph}.  Moreover, the distance to
\object{U Sco} is much more far away ($6-14$~kpc) than that to \object{RS Oph}
($0.6-1.6$~kpc).  If the estimated column densities both for \object{RS Oph}
and \object{U Sco} are correct, we expect a luminous supersoft X-ray
phase also in RS Oph.

However, the cool giant's wind may obscure supersoft X-rays.
\citet{ori93} observed
\object{RS Oph} with {\it ROSAT} in the quiescent phase 7 yrs
after the outburst.  The observed flux 
of $(3-20) \times 10^{31}$erg~s$^{-1}$
is too low to be compatible with the accretion luminosity given in
equation (\ref{accretion_X_eq}).
One possible explanation is an absorption by the cool giant's wind
as suggested by \citet{anu99}.

If the cool wind obscures the WD,
supersoft X-rays of nuclear burning may not be
observable until the cool giant's wind is swept away.
In the 1985 outburst, X-ray rapidly decayed from day 70--80.
\citet{mas87} interpreted this decay as that the blast shock had broken
out of the cool giant's wind on day $\sim 70$.  We regard 
day 70--80 as the time when the cool giant's wind is swept away.
If the WD wind stops before day 70--80, accurate turn-on time of
supersoft X-ray may be missed.  However, we are still able
to determine the WD mass only from the turn-off time if it is detected.

The cool giant's wind, once swept away by the hot WD ejecta,
will again fill its Roche lobe and accrete onto the hot component
in a timescale of
$\Delta t = a / v \sim 300 R_\sun / 10 \mbox{~km~s}^{-1}
= 300$ days.  When accretion onto the WD resumes, the accretion disk
brightens up, which is roughly consistent with the recovery
of the quiescent $V$ magnitude from 12 to 11 mag on day $\sim 400$ of
the previous outburst \citep[see, e.g., Fig.1 of][]{eva88}.
Once the cool wind covers the hot WD, supersoft X-rays are absorbed.
Therefore, a luminous supersoft X-ray phase may be observable only
during day $\sim 80-400$. 

Another good indicator of WD mass is a UV 1455 \AA\  band of 
continuum flux proposed by \citet{cas02}.
\citet{hac05k} and \citet{kat05h} showed
that their optically thick wind model well reproduces simultaneously
both the supersoft X-ray and UV 1455 \AA\  light curves of
the classical nova V1974 Cyg.  They estimated the WD mass as
$\sim 1.05 ~M_\sun$.  Thus we are able to determine the WD mass
using the UV 1455 \AA\  light curve when the chemical composition
of ejecta is accurately given.  Determination of chemical
composition is definitely required for \object{RS Oph}. 


%% Included in this acknowledgments section are examples of the
%% AASTeX hypertext markup commands. Use \url without the optional [HREF]
%% argument when you want to print the url directly in the text. Otherwise,
%% use either \url or \anchor, with the HREF as the first argument and the
%% text to be printed in the second.

\acknowledgments

We thank VSOLJ and AAVSO for the visual data on \object{RS Oph}.
We are also grateful to the anonymous referee
for useful comments to improve the manuscript.
This research has been supported in part by the Grant-in-Aid for
Scientific Research (16540211, 16540219)
of the Japan Society for the Promotion of Science.

\end{document}